\documentclass[11pt]{article}
\usepackage{geometry}                
\usepackage{graphicx}
\usepackage{amsmath, amsthm, amsfonts}
\usepackage{amssymb}

\newcommand{\sect}[1]{\setcounter{equation}{0}\section{#1}}

\usepackage{color}

\usepackage{epstopdf}
\title{The general Racah algebra as the symmetry algebra\\
of  generic systems on  pseudo--spheres}
\author{
\c{S}. Kuru$^a$, I. Marquette$^b$ and  J. Negro$^c$ 
\bigskip
\\
\noindent
$^a$Department of Physics, Faculty of Science, Ankara
University, 06100 Ankara, Turkey
\\ 
\noindent
$^b$School of Mathematics and Physics, The University of Queensland
Brisbane, \\ QLD 4072, Australia
\\ 
 \noindent
$^c$Departamento de F\'{\i}sica Te\'orica, At\'omica y
\'Optica, Universidad de Valladolid,\\  47011 Valladolid, Spain
\\~~E-mail:   jnegro@fta.uva.es
}

\begin{document}
\maketitle

\begin{abstract}
\noindent 
We characterize the symmetry algebra of the generic superintegrable system on a pseudo-sphere
corresponding to the homogeneous space $SO(p,q+1)/SO(p,q)$ where $p+q={\cal N}$, ${\cal N}\in\mathbb N$.
We show that this algebra is independent of the signature $(p,q+1)$ of the metric
and that it is the same as the Racah algebra ${\cal R}({\cal N}+1)$. The spectrum obtained
from ${\cal R}({\cal N}+1)$ via the Daskaloyannis method depends on undetermined signs that can be associated to the signatures. Two examples are 
worked out explicitly for the cases $SO(2,1)/SO(2)$ and $SO(3)/SO(2)$ where it is shown that their spectrum obtained by means of separation of variables coincide with 
 particular choices of the signs corresponding to the specific signatures of the spectrum for the symmetry algebra 
 ${\cal R}(3)$.
\end{abstract}
%
%
%
%
%

\sect{Introduction}

The Racah algebra ${\cal R}(3)$ has been  applied to describe the recoupling  of three copies of $su(1,1)$, but it has been used in many other contexts, for instance it was shown that it is the symmetry algebra of the so called ``generic superintegrable system'' on the sphere ${\cal S}^2$ \cite{vinet14,vinet19,post07,post11}. Another
related property is that this algebra can be identified as the commutant of 
$o(2)\oplus o(2) \oplus o(2)$ in $o(6)$. All these properties can be extended
to ${\cal N}$ copies of $su(1,1)$ and then it is called the generalized ${\cal R}({\cal N})$
Racah algebra \cite{vinet19}. The connection with the Howe duality and embedding into Bannai-Ito algebra was also discussed. The Racah algebra ${\cal R}(3)$ which is included in Askey-Wilson QAW(3) algebra \cite{gra92} has been also applied in position dependent mass systems \cite{que07}. 

Along this work we will consider the symmetry algebra of generic systems defined 
on a pseudo--sphere in an ambient space $\mathbb R^{p+q+1}$ endowed with a pseudo Riemannian
metric $g^{\mu\nu}$ with signature $(p,q+1)$. Other notations can be used, but in order to fix a kind of systems to be considered later let us define a pseudo-sphere ${\cal S}^{p,q}$  through the equation
\begin{equation}\label{orbit}
{s_1}^2+\dots + {s_p}^2- {s_{p+1}}^2-\dots - {s_{p+q+1}}^2= - 1,\qquad {\bf s}\in \mathbb R^{p+q+1}\,.
\end{equation}
This is an orbit of the pseudo orthogonal group $SO(p,q+1)$ in the ambient space 
${\mathbb R}^{p+q+1}$,  and it can also be seen as the homogeneous space ${\cal S}^{p,q} = SO(p,q+1)/SO(p,q)$. We shall see that all the symmetry algebras of the ``generic systems'' associated to any such metric
$g^{\mu\nu}$ with $p+q={\cal N}$ fixed, will coincide. This means
that their symmetry algebras can be identified to the same Racah algebra ${\cal R}({\cal N}+1)$. This algebra is  the commutant of $\oplus^{{\cal N}+1} o(2) $ in the Lie algebra $o(2p,2q+2)$ \cite{olmo08,calzada11}, which should be the isomorphic to the commutant in $so(2{\cal N}+2)$, i.e., independent of the signature.

We will analyse  the possible discrete spectra for the particular case
${\cal R}(3)$, by following the method of Daskaloyannis \cite{das01}. Then, we will show that the formulas so obtained include the two cases which
have  discrete spectrum corresponding to the systems defined on
the sphere ${\cal S}^2\equiv {\cal S}^{0,2} \approx SO(3)/SO(2)$ and on the (two-sheeted) hyperbolic space 
${\cal H}^2\equiv {\cal S}^{2,0} \approx SO(2,1)/SO(2)$. The signature
of the initial Hamiltonian can be identified in the final formula of the spectrum.

The organization of the paper is as follows. In Section 2, we define this kind of generic superintegrable systems on a general pseudo--sphere. We supply the form of the quadratic symmetries
and the symmetry algebra which is independent of the metric coefficients $g^{\mu\nu}$. It is also included the quantum coefficient $\hbar$ in all the terms
so that the classical limit of the system is obtained by taking $\hbar\to 0$.
Next, in Section~3, the symmetry algebra for ${\cal N}=3$ is written in the form of a Daskaloyannis
algebra \cite{das06}. In this way we have computed the possible
discrete spectrum of the symmetry algebra. In Section 4,  we supply the spectrum of the quantum systems defined
on the sphere ${\cal S}^2$ \cite{olmo06} and on the hyperboloid ${\cal H}^2$
\cite{olmo08}, which can be obtained by means of separation of variables. We will check that, indeed, these formulas are included in the ones obtained in Section~3. The paper will end with some remarks and conclusions in Section~4.

\sect{Generalized Racah Algebra}

Let the pseudo--sphere
${\cal S}^{p,q}$ be defined as the surface
\begin{equation}\label{orbit2}
g_{ij}s^is^j= -1,\qquad (s^1,\dots, s^{{\cal N}+1})\in {\mathbb R}^{{\cal N}+1}
\end{equation} 
where the $g_{ik}$'s are metric coefficients $g=(g_{ik})$ with signature $(p,q+1)$ as mentioned above.
Then, the Hamiltonian of a ${\cal N}= p+q$ dimensional superintegrable system on 
${\cal S}^{p,q}$
is defined by
\begin{equation}\label{hg}
H=\frac{1}{2}\, g_{ik} g_{jl} J_{ij}J_{kl} +  g_{ii} \frac{a_{i}}{s_{i}^{2}},\qquad
a_i\in \mathbb{R}
\end{equation}
where 
\begin{equation}\label{jg}
J_{ij}= \hbar g_{jk}  s_{i} \partial_{k}- \hbar g_{ik} s_{j} \partial_{k}\,
\end{equation}
are the anti-Hermitian generators of the Lie algebra $SO(p,q+1)$ which leaves invariant the
pseudo--sphere. We have also included the quantum constant $\hbar$ in order
to consider later the classical limit.
We use the convention of sum in the repeated indexes with some care: if one of the repeated indexes 
is in both sides of the equation and it is taking part of the definition of a component
of a tensor, it will not be summed.

The symmetries for the generic Euclidean case have been known from some time ago \cite{Miller}.  In the case of generic metric $g^{\mu\nu}$ (see Ref.~\cite{calzada11} were the signature was considered) the second order symmetries of the above Hamiltonian (\ref{hg}) have the form
\begin{equation}\label{qg}
Q_{ij}=- g_{ik}g_{jl} J_{ij} J_{kl} + g_{ii} g_{jj} a_{i} \frac{s_{j}^{2}}{s_{i}^{2}}+ g_{jj} g_{ii} a_{j} \frac{s_{i}^{2}}{s_{j}^{2}}\,.
\end{equation}
The commutations of these symmetries are as follows. Firstly, the commutators 
\begin{equation}\label{qqc}
[ Q_{ij},Q_{ik}]= \hbar\, C_{ijk}\,,
\end{equation}
will lead to the third order symmetry operators $C_{ijk}$. All these operators 
$\{ Q_{ij}, C_{klm}\}$ will close a quadratic symmetry algebra:
\begin{equation}\label{sa}
\begin{array}{l}
[ Q_{jk,},C_{ijk}]= \hbar\Big(8   Q_{ik}Q_{jk} - 8   Q_{jk} Q_{ij} + 8(-\hbar^{2} +2 a_{j}  )Q_{ik}
\\[2ex]
\hskip2cm - 8 (-\hbar^{2} + 2 a_{k} ) Q_{ij} + 8 (a_{j}-a_{k})\hbar^{2}\Big)\,, 
\\[2ex]
[ Q_{kl}, C_{ijk}]= \hbar\Big( 8   Q_{ik} Q_{jl} - 8   Q_{il} Q_{jk}+ 
4 \hbar^{2} Q_{ik} + 4 \hbar^{2} Q_{jl} - 4 \hbar^{2} Q_{il}  - 4 \hbar^{2} Q_{jk}\Big)\,,
\\[2ex]
[ C_{ijk}, C_{jkl}]=\hbar\Big(-8   C_{jkl} Q_{ij}  - 8   C_{ikl} Q_{jk} - 8   C_{ijk} Q_{jl}
\\[2ex]
\hskip2cm + 4  \hbar^{2} C_{jkl}  - 4  \hbar^{2} C_{ijk}  - 8 (-\hbar^{2} + 2 a_{j} ) C_{ikl } \Big)\, ,
\\[2ex]
[C_{ijk}, C_{klm}]= \hbar\Big(- 8   C_{ilm} Q_{jk} - 8   Q_{ik} C_{jlm}
 + 4 \hbar^{2} C_{ilm} - 4 \hbar^{2} C_{jlm}\big) \,,
 \end{array}
\end{equation}
\[ [ C_{ijk}, C_{lmn}]=0 \,. 
\]
These symmetry operators are invariant under cyclic permutation of their subindexes,
\[ 
Q_{ij}=Q_{ji}, \qquad C_{ijk}= C_{kij} = C_{jki} (= - C_{jik})\,. 
\]
Remark that the Hamiltonian can be expresed in terms of the quadratic symmetries $Q_{ij}$:
\begin{equation}\label{qh}
\sum_{i<j} \alpha_{ij} Q_{ij} -\alpha_{0} H - \alpha_{00} =0 
\end{equation}
where
\[ \alpha_{ij} = \frac{n(n+1)}{2} \alpha_{0},\qquad\alpha_{00}=- \sum_{i}^{n} a_{i} \alpha_{0}\, .
\]

Therefore, we see that the signature in the initial Hamiltonian determine the constants of
motion (\ref{qg}), but the symmetry algebra (\ref{sa}) does not include any track of the metric coefficients 
$g^{ij}$, hence it is the same for any generic system (\ref{hg}) on a pseudo Riemannian surface.

\subsection{Classical case}

In the previous section the quantum symmetry algebra has terms with the quantum
constant $\hbar$. In order to get the symmetry algebra of the corresponding 
classical system (replacing
the quantum by classical magnitudes in the initial Hamiltonian) 
it is enough to take the limit $\hbar \to 0$ and
replace quantum commutators $[A,B]=i\hbar C$ by Poisson brackets (PB): $\{A,B\}_{PB}= C$. In this
way,  we get the following algebra:
\begin{equation}\label{cs}
\{ Q_{ij},Q_{ik}\}_{PB}= C_{ijk}\,,
\end{equation}
\begin{equation}
\begin{array}{l}
\{ Q_{jk,},C_{ijk}\}_{PB}= -8 Q_{ik}Q_{jk} + 8 Q_{jk} Q_{ij} - 16 a_{j} Q_{ik}
+ 16 a_{k} Q_{ij}\, ,
\\[2ex]
\{ Q_{kl}, C_{ijk}\}_{PB}= -8 Q_{ik} Q_{jl} + 8 Q_{il} Q_{jk} \, ,
\\[2ex]
\{ C_{ijk}, C_{jkl}\}_{PB}=8  C_{jkl} Q_{ij}  + 8 C_{ikl} Q_{jk} + 8  C_{ijk} Q_{jl}
 + 16 a_{j} C_{ikl }\, ,
\\[2ex]
\{C_{ijk}, C_{klm}\}_{PB}= 8  C_{ilm} Q_{jk} + 8 Q_{ik} C_{jlm} \, ,
\\[2ex]
\{ C_{ijk}, C_{lmn}\}_{PB}=0 \,.
\end{array} 
\end{equation}
The symmetries satisfy the same cyclic relations as in the quantum  case:
\[ Q_{ij}=Q_{ji},\qquad C_{ijk}= C_{kij} = C_{jki}\,. \]

If we compare the classical and the quantum symmetry algebras, we can appreciate that, as usual,  the classical one is simpler because some of the terms in the quantum commutators vanish in the classical limit.

\sect{An Example: The Three--Dimensional Case}

The set of  of second order symmetries in the three-dimensional case are $Q_{12}$, $Q_{13}$, $Q_{23}$. As the system is superintegrable there must be three independent symmetries including the Hamiltonian. 
Thus,  we can choose two of them $Q_{12}$, $Q_{13}$ together with $H$ as the independent set.
The other one $Q_{23}$, with the help of (\ref{qh}), can be expressed as 
\[
Q_{23}=-Q_{12}-Q_{13}+\frac{1}{3}H-\frac{1}{3}\sum a_{i}\,.
\] 

The only third order symmetry of the type $C_{ijk}$ is $C_{213}$. 
In this section we will take $\hbar=1$ since we will not consider the classical counterpart. 
Then, the set of the three symmetries
$\{Q_{12},Q_{13},C_{213}\}$ close, together with (\ref{qqc}), the following algebra,

\begin{equation}
\begin{array}{l}
[Q_{12},C_{213}]= +8 \{Q_{12},Q_{13}\}+ 16(-1+a_{1}+a_{2}) Q_{13} 
\\[1.5ex]
\hskip1.5cm + 4 (-2 +6 a_{1}+2a_{2}+2a_{3}) Q_{12} - 8 (-1 +2 a_{1})H
\\[1.5ex]
\hskip1.5cm - 8 H Q_{12}+ 8 Q_{12}^{2} 
+4 ( -4 a_{1}+4 a_{1}^{2}+4 a_{1}a_{2}- 2 a_{3}+4 a_{1} a_{3} )  \,,
\\[1.5ex]
[Q_{13},C_{213}]=- 8 \{Q_{12},Q_{13}\} - 8 Q_{13}^{2}+ 8 H Q_{13}
\\[1.5ex]
\hskip1.5cm- 4 (-2 + 6 a_{1}+ 2 a_{2}+ 2 a_{3}) Q_{13}-16 (-1 +a_{1}+a_{3})Q_{12}
\\[1.5ex]
\hskip1.5cm + 8 ( -1 +2 a_{1})H - 4( -4 a_{1}+4 a_{1}^{2}- 2 a_{2}+4 a_{1} a_{2}+ 4 a_{1}a_{3} )  \, .
\end{array}
\end{equation}
This algebra can be rewritten in the form of a Daskaloyannis type algebra 
\cite{das01,das06} spanned by the generators $\{A,B,C\}$ and having the commutations in the form
\begin{equation}
\begin{array}{c}
[A,B]=C \,,
\\[1.5ex]
 [ A,C]=\alpha A^{2} + \gamma \{A,B\} +\delta A+ \epsilon B + \zeta  \,,
 \\[1.5ex]
 [B,C]= a A^{2}- \gamma B^{2} - \alpha \{A,B\} + d A - \delta B + z  \,,
\end{array}
\end{equation}
where the structure constants in our case take the values:
\[ 
\alpha=8,\quad \gamma=8,\quad
\epsilon = 16,\quad a=0,\quad \delta= 4 (-2 + 6 a_{1}+ 2 a_{2}+ 2 a_{3}) - 8 H ,
\]
\[
d=-16(-1+a_{1}+a_{3}),\quad \zeta= 4 (- 4 a_{1} + 4 a_{1}^{2} + 4 a_{1} a_{2} - 2 a_{3} + 4 a_{1} a_{3}) - 8 (-1 +2 a_{1}) H, 
\]
\[ 
z= 8 (-1 +2a_{1})H - 4 (-4 a_{1} + 4 a_{1}^{2} -2 a_{2} +4 a_{1}a_{2} + 4 a_{1}a_{3} ) \,.
\]
The Casimir of this algebra is given by
\begin{equation}
K=C^{2}-\alpha \{A^{2},B\} - \gamma\{A,B^{2}\}+( \alpha, \gamma -\delta) \{A,B\}
\end{equation}
\[ + (\gamma^2 -\epsilon ) B^{2} + ( \gamma \delta -2 \zeta ) B + \frac{2 a}{3} A^{3} 
+ ( d+ \frac{a \gamma}{3} +\alpha^{2})A^{2} + (\frac{a\epsilon}{3} + \alpha \delta +2 z) A \,,
\]
which can be written in the present realization in terms of the Hamiltonian:
\begin{equation}
\begin{array}{l}
K=4(-3 +4 a_{1})H^{2} - 8( 6 - 21 a_{1}+4 a_{1}^{2}-3 a_{2}+4 a_{1}a_{2}-3 a_{3}+4 a_{1} a_{3})H 
\\[1.5ex]
\hskip1.5cm + 4 \big( 20 a_{1}-39 a_{1}^{2}+4 a_{1}^{3} + 4 a_{2}- 30 a_{1}a_{2} 
\\[1.5ex]
\hskip1.5cm
+8 a_{1}^{2} a_{2}-3 a_{2}^{2} + 4 a_{1} a_{2}^{2} + 4 a_{3}-30 a_{1} a_{3}+ 8 a_{1}^{2}a_{3}
\\[1.5ex]
\hskip1.5cm
+ 6 a_{2}a_{3}-8 a_{1}a_{2}a_{3}-3a_{3}^{2}+4 a_{1}a_{3}^{2} \big)\,.
\end{array}
\end{equation}

This formula, providing a link between the Casimir and the Hamiltonian, will allow to establish the realization of the quadratic algebra as a deformed oscillator algebra \cite{das01} of the form
\[  [N,b]=-b,\quad [N,b^{\dagger}]=b^{\dagger},\quad bb^{\dagger}=\Phi(N+1),\quad b^{\dagger}b=\Phi(N) 
\]
where $\Phi(N)$ is the structure function which is a polynomial in terms of the number operator $N$ and the representation dependent parameter $u$.
Remark that this algebraic approach has been extended to polynomial algebras \cite{mar14} with three generators and applied to higher rank quadratic algebras \cite{mar18}. The expression of $\Phi$ is
\[
\Phi (N)=768 \left(\alpha  \epsilon ^2+4 \gamma ^2 \zeta -2 \gamma  \delta  \epsilon \right)^2+32 \gamma ^4 (2 (N+u)-1)^2 \]
\[ \left(12 (N+u)^2-12 (N+u)-1\right) \left(3 \alpha ^2 \epsilon ^2+4 \alpha  \gamma ^2 \zeta -6 \alpha  \gamma  \delta  \epsilon +2 a \gamma  \epsilon ^2\right.\]
\[\left.+2 \gamma ^2 \delta ^2-4 \gamma ^2 d \epsilon +8 \gamma ^3 z-48 \gamma ^6\right) (2 (N+u)-3) (2 (N+u)-1)^4 (2 (N+u)+1) \left(\alpha ^2 \epsilon -\alpha  \gamma  \delta \right.\]
\[\left. +a \gamma  \epsilon -\gamma ^2 d\right)-256 \gamma ^2 (2 (N+u)-1)^2 \left(3 \alpha ^2 \epsilon ^3+4 \alpha  \gamma ^4 \zeta \right.\]
\[+12 \alpha  \gamma ^2 \zeta  \epsilon -9 \alpha  \gamma  \delta  \epsilon ^2+a \gamma  \epsilon ^3+2 \gamma ^4 \delta ^2\]
\[-12 \gamma ^3 \delta  \zeta +6 \gamma ^2 \delta ^2 \epsilon +2 \gamma ^4 d \epsilon -3 \gamma ^2 d \epsilon ^2-4 \gamma ^5 z+\]
\[\left.12 \gamma ^3 z \epsilon \right)+\gamma ^8 (2 (N+u)-3)^2 (2 (N+u)-1)^4 (2 (N+u)+1)^2 \left(3 \alpha ^2+4 a \gamma \right)-3072 \gamma ^6  K(2 (N+u)-1)^2\,.\]
Using the structure constants and the Casimir operator as expressed in terms of the central element of the algebra (the Hamiltonian $H$) one get

\begin{equation}
\Phi(N,u,E)= 3221225472 ( a_{1}^{2}+a_{2}^{2}-2 a_{1}( a_{2}+(1-2(N+u))^{2})
\end{equation}
\[-2 a_{2}(1-2(N+u))^{2}+4(1-2(N+u))^{2}(-1+(N+u))(N+u) \]
\[ (a_{3}^{2}+E^{2} -2 a_{3} (E+(1-2(N+u))^{2}) \]
\[-2E(1-2N)^{2}+4 (1-2(N+u))^{2}(-1+(N+u))(N+u))\,.\]

The structure function in this form allow us to characterize the finite dimensional unitary representations and in this way it will lead to the discrete spectrum for the energy $E$. 
This formula is a polynomial of degree 8 in the number operator $N$, but it is not yet in a convenient form. Using the parameter
\[  m_{i}^{2}= 1+ 4 a_{i},\quad i=1,2,3 \]
and introducing 
\[  - \tilde{E}^{2}=-1+4E \,,\]
we can reexpress the structure function as
\begin{equation}
\Phi(N,u,E)= 824633720832 ( N+u-N_{1})( N+u-N_{2})
\end{equation}
\[( N+u-N_{3})( N+u-N_{4})( N+u-N_{5})\]
\[( N+u-N_{6})( N+u-N_{7})( N+u-N_{8})\]
with
\[\begin{array}{ll}
N_{1}=\frac{1}{4}( 2- ( m_{1}-m_{2})),\qquad & N_{5}=\frac{1}{4}( 2- ( \tilde{E}-m_{3})),
 \\[1.5ex]
N_{2}=\frac{1}{4}( 2+ ( m_{1}-m_{2})), \qquad & N_{6}=\frac{1}{4}( 2+ ( \tilde{E}-m_{3})),
\\[1.5ex] 
N_{3}=\frac{1}{4}( 2- ( m_{1}+m_{2})), \qquad & N_{7}=\frac{1}{4}( 2- ( \tilde{E}+m_{3})),
\\[1.5ex]
N_{4}=\frac{1}{4}( 2+ ( m_{1}+m_{2})), \qquad & N_{8}=\frac{1}{4}( 2+ ( \tilde{E}+m_{3}))\,.
\end{array}\]

This is a factorized form that will facilitate greatly the study of the finite dimensional unitary representations. The constraints that need to be satisfied in order to get finite dimensional unitary representations are the following:
\[ \Phi(0,u,E)=0,\quad \Phi(p+1,u,E)=0,\quad \Phi(\nu,u,E)>0 \qquad \forall\, \nu=1,...,p \,. \]
The first condition $\Phi(0,u,E)=0$ provide
\[ u=\frac{1}{4}(2+m_{1} \epsilon_{1}+m_{2} \epsilon_{2}) \]
where the parameters $\epsilon_{1}=\pm1,\epsilon_{2}=\pm 1$ allow to describe the different solutions in a unified way. They supply us with the structure function under the form
\[ \Phi(N,E)= 12582912 (-4N + m_{1}(-1-\epsilon_{1})+m_{2}(-1-\epsilon_{2})
(-4N + m_{1}(1-\epsilon_{1})+m_{2}(-1-\epsilon_{2}) \]
\[ (-4N + m_{1}(-1-\epsilon_{1})+m_{2}(1-\epsilon_{2})(-4N + m_{1}(1-\epsilon_{1})+m_{2}(1-\epsilon_{2})\]
\[  (-\tilde{E}-4N -m_{3}+m_{1} \epsilon_{1} - m_{2} \epsilon_{2} )
 (\tilde{E}-4N -m_{3}+m_{1} \epsilon_{1} - m_{2} \epsilon_{2} )\]
\[  (-\tilde{E}-4N m_{3}+m_{1} \epsilon_{1} - m_{2} \epsilon_{2} )
 (\tilde{E}-4N m_{3}-m_{1} \epsilon_{1} - m_{2} \epsilon_{2} )\,.\]
The second condition $\Phi(p+1,u,E)=0$ provides

\[ \tilde{E}=4(p+1)- \epsilon_{3} m_{3} - m_{2} \epsilon_{2} - m_{1} \epsilon_{1} \]
and
\[ E= \frac{1}{4} - \frac{1}{4}( \epsilon_{3} m_{3} + \epsilon_{2} m_{2} + \epsilon_{1} m_{1}- 4(p+1) )^{2}\,. \]
We will introduce the new parameters $l_i$ whose meaning will be explained below:
\[
m_i = 2\, l_i,\quad i=1,2,3\,.
\]
Then, the spectrum will take the form
\begin{equation}\label{e} E= \frac{1}{4} - (  \epsilon_{3} l_{3} + \epsilon_{2} l_{2} + \epsilon_{1} l_{1}- 2(p+1) )^{2} 
\end{equation}
where $a_i = l_i^2-1/4$. The degeneracy of each energy level, determined
by $p$, is $p+1$.

\subsection{The spectrum of the three dimensional system}

We will consider an example corresponding to metric coefficients of
different signs: ($g_{11}=+1,  g_{22}=+1,  g_{33}=-1$).
The surface is the hyperbolic space ${\cal H}^2$, the upper component of the two dimensional two-sheeted hyperboloid given by 
$s_1^2+s_2^2-s_3^2=-1$, inside the real space ${\mathbb R}^3$.

The superintegrable generic Hamiltonian in this case has the form \cite{olmo08}:
\begin{equation}\label{hhc3}
H =-J_{1}^2-J_{2}^2+J_{3}^2+\frac{\ell_1^2-\frac{1}{4}}{s_1^2}+\frac{\ell_2^2-\frac{1}{4}}{s_2^2}-\frac{\ell_3^2-\frac{1}{4}}{s_3^2}\,,
\end{equation}
where $\ell=(\ell_1,\ell_2,\ell_3)\in \mathbb R^3$, 
and the anti-Hermitian generators of
$SO(2,1)$ are
\begin{equation}\label{j3}
\begin{array}{cc}
J_{1}=s_2\,\partial_{s_3}+s_3\,\partial_{s_2}\,,
\\[2.ex]
J_{2}=s_3\,\partial_{s_1}+s_1\,\partial_{s_3}\,,
\\[2.ex]
J_{3}=-s_1\,\partial_{s_2}+s_2\,\partial_{s_1}\,.
\end{array}
\end{equation}
Here, $J_{3}$ generates true rotations around $s_3$, while $J_{1}, J_{2}$ are generators of the pseudo-rotations around $s_1$ and $s_2$, respectively. In terms of these generators the kinetic part of the Hamiltonian (\ref{hhc3}) is (proportional to) the $so(2,1)$ Casimir operator
\begin{equation}\label{c3}
C=J_{1}^2+J_{2}^2-J_{3}^2\,.
\end{equation}

We can parametrize  the hyperbolic surface in the following way \cite{olmo08}:
\begin{equation}\label{p3}
 s_1=\sinh \xi\, \cos \theta\,,\qquad s_2=\sinh \xi\, \sin \theta\,,\qquad
 s_3=\cosh \xi\,,
\end{equation}
where $0\leq\theta<2 \pi,\,0\leq \xi<\infty$. Using this parametrization, the Hamiltonian (\ref{hhc3}), takes the form
\begin{equation}\label{h3p}
H =-\partial_{\xi}^2-\coth\xi\,\partial_{\xi}
-\frac{\ell_3^2-\frac{1}{4}}{\cosh{\xi}^2}
+\frac{1}{\sinh{\xi}^2}
\left( -\partial_{\theta}^2+\frac{\ell_2^2-\frac{1}{4}}{\cos{\theta}^2}
+\frac{\ell_1^2-\frac{1}{4}}{\sin{\theta}^2}\right)\,.
\end{equation}
The corresponding eigenvalue equation is $H\, \Psi(\xi,\theta)=E\, \Psi(\xi,\theta)$. 

It can be shown by separating variables \cite{olmo08} that the discrete spectrum is given by
\begin{equation}\label{h3p2e}
E = \frac{1}{4} -(\ell_3-\ell_1-\ell_2-2(n+m+1))^2\,,
\end{equation}
where $m,n\in\mathbb Z^+$ and $\ell_3-\ell_1-\ell_2-2(n+m+1)>0$. The degeneracy
is given by the $P+1$ values of $m,n$ such that $m+n= P$, with $0\leq P <(\ell_3-\ell_1-\ell_2)/2 -1$. 

We can compare this formula with that obtained following the Daskaloyannis method  
given by (\ref{e}). 
We see that they will coincide if we make the identifications $g_{ii} = \epsilon_i$ and $p =n+m\equiv P$. The values of $p=0,1\dots$
are subject to the condition
\begin{equation}\label{h3p2m}
\dfrac{\ell_3-\ell_1-\ell_2}{2}-1>p\geq 0 \,.
\end{equation}
The allowed values of $p$ determine the finite spectrum and the degeneracy of each level is
$p+1$. Only the ground level $p=0$ will be a singlet.

Let us briefly mention the case with metric coefficients of
equal signs: $g_{ii}=1$, $i=1,2,3$.
The surface is ${\cal S}^2$, the two dimensional sphere,
$s_1^2+s_2^2+ s_3^2=1$ in ${\mathbb R}^3$ (equivalently we could take $g_{ii}=-1$
and the equation of the sphere with an overall $-1$ sign).
The generic Hamiltonian in this case has the usual form \cite{olmo06}:
\begin{equation}\label{hhc4}
H =-(J_{1}^2+J_{2}^2+J_{3}^2)+\frac{\ell_1^2-\frac{1}{4}}{s_1^2}+\frac{\ell_2^2-\frac{1}{4}}{s_2^2}+\frac{\ell_3^2-\frac{1}{4}}{s_3^2}\,,
\end{equation}
where $\ell=(\ell_1,\ell_2,\ell_3)\in \mathbb R^3$, 
and the anti-Hermitian generators of $SO(3)$.
The spectrum is given by
\begin{equation}\label{h3p2e}
-E = \frac{1}{4} -(\ell_1+\ell_2+\ell_3+2(n+m+1))^2\,,
\end{equation}
where $m,n\in\mathbb Z^+$. There is an infinite number of discrete energy levels whose degeneracy also is given by the $P+1$ values of $m,n$ such that $m+n= P$. 
The formula (\ref{e}) applies also here provided we take $\epsilon_i=-1$, $i=1,2,3$,
as well as an overall change of sign coming from the initial Hamiltonian (\ref{hhc4}).

\sect{Conclusions}

We derived the symmetry algebra of a generic Hamiltonian on the pseudo--sphere 
${\cal S}^{p,q}$ corresponding to an ambient space $\mathbb R^{p+q+1}$ with metric $g^{\mu\nu}$ of signature $(p,q+1)$. We showed that this algebra is the same, independently of the metric, as the general Racah algebra ${\cal R}({\cal N}+1)$, where ${\cal N}=p+q$. 

We considered in detail the particular case of the case $g^{\mu\nu}={\rm diag}(1,1,-1)$
and the homogeneous space $SO(2,1)/SO(2)$ of the two sheeted hyperbolic space and 
the corresponding generic Hamiltonian. We constructed the symmetry algebra, its Casimir operator, the realization as a deformed oscillator algebra and calculated the energy spectrum algebraically which depend on some signs $\epsilon_i$. We compared this algebraic spectrum  with the ``physical spectrum'' obtained via separation of variables of the corresponding Schr\"odinger equation and showed how they both coincide for a choice of the signs which is given by the signature of the metric.
In Ref.~\cite{olmo08} it was shown that the symmetry algebra of each of the generic Hamiltonians in $SO(2,1)/SO(2)$ can be identified
as the commutant of $\oplus^3so(2)$ in the enveloping algebra of $so(4,2)$. Meanwhile, in the standard case of the generic system on $SO(3)/SO(2)$ it is known, based on the Howe duality, that the symmetry algebra is identified
as the commutant of $\oplus^3o(2)$ in the enveloping algebra of $o(6)$. It seems as
if the symmetry operators, being quadratic,  loose the track of the signature and they
have the same algebraic structure for any metric signature. There are even some homogeneous spaces,
for instance $SO(2,1)/SO(1,1)$, where the inner product, in the space of wave functions on the manifold, is not positive definite and
therefore there is not the notion of bound states. In this case the algebra in principle would be the same as in the above cases because the homogeneous space is not used in the derivation of the algebra. The discrete spectrum obtained by Daskaloyannis method, however, would give
the same formula although no discrete spectrum is available in this case.

 \section*{Acknowledgments}

This work was partially supported by Junta de Castilla y Le\'on (BU229P18, VA137G18).  \c{S}.~Kuru acknowledges Ankara University and the warm hospitality at Department
of Theoretical Physics, University of Valladolid, where part of this work has been done. I.~Marquette was supported by Australian Research Council with a Future Fellowship FT180100099.


\end{document}